\documentclass[letterpaper, 10 pt, conference]{ieeeconf}  

\IEEEoverridecommandlockouts                              
 
\usepackage{graphicx} 
\usepackage{amsmath} 
\usepackage{amssymb}  
\usepackage{caption}
\usepackage{url,times,booktabs, tabularx}
\usepackage[colorlinks=false,pdfborder={0 0 0}]{hyperref}
\usepackage{units}
\usepackage{multirow,array}
\usepackage{balance}
\usepackage{float}
\usepackage{soul}
\usepackage{color}
\usepackage{cleveref}
\usepackage{wrapfig}
\usepackage{booktabs}
\usepackage[activate]{microtype}
\usepackage{array}
\usepackage{bm}
\usepackage{bbm}
\newcolumntype{P}[1]{>{\centering\arraybackslash}p{#1}}
\newcommand{\eg}{e.\,g.,\ }
\newcommand{\ie}{i.\,e.,\ }

\newcolumntype{R}[1]{>{\raggedleft\let\newline\\\arraybackslash\hspace{0pt}}m{#1}}

\usepackage{times}
\usepackage[T1]{fontenc}
\usepackage{pslatex}
\usepackage{dsfont}
\usepackage{fancyhdr}
\title{\LARGE \bf
End-to-end Acoustic-linguistic Emotion and Intent Recognition \\Enhanced by Semi-supervised Learning
}

\author{Zhao Ren$^{1}$, Rathi Adarshi Rammohan$^{1}$, Kevin Scheck$^{1}$, Sheng Li$^{2}$, Tanja Schultz$^{1}$
\thanks{This study is supported by the Deutsche Forschungsgemeinschaft (DFG, German Research Foundation) through the projects ``Silent Paralinguistics" (40301193) and ``MyVoice: Myoelectric Vocal Interaction and Communication Engine'' (460884988).}
\thanks{$^{1}$Z. Ren, R. Rammohan, K. Scheck, and T. Schultz are with the Cognitive Systems Lab, University of Bremen, Germany ({\tt\small zren@uni-bremen.de}).}%
\thanks{$^{2}$S. Li is with the Institute of Science Tokyo, Japan.}
} 

\begin{document}

\maketitle
\thispagestyle{fancy}         
\fancyhead{}                     
\lhead{\scriptsize This work has been submitted to the IEEE for possible publication. Copyright may be transferred without notice, after which this version may no longer be accessible.}       \chead{}
\lfoot{}
\cfoot{\thepage}  
\rfoot{}
\renewcommand{\headrulewidth}{0pt}  
\renewcommand{\footrulewidth}{0pt}
\pagestyle{empty}

\begin{abstract}
Emotion and intent recognition from speech is essential and has been widely investigated in human-computer interaction. The rapid development of social media platforms, chatbots, and other technologies has led to a large volume of speech data streaming from users. Nevertheless, annotating such data manually is expensive, making it challenging to train machine learning models for recognition purposes. To this end, we propose applying semi-supervised learning to incorporate a large scale of unlabelled data alongside a relatively smaller set of labelled data. We train end-to-end acoustic and linguistic models, each employing multi-task learning for emotion and intent recognition. Two semi-supervised learning approaches, including fix-match learning and full-match learning, are compared. The experimental results demonstrate that the semi-supervised learning approaches improve model performance in speech emotion and intent recognition from both acoustic and text data. The late fusion of the best models outperforms the acoustic and text baselines by joint recognition balance metrics of 12.3\,\% and 10.4\,\%, respectively.


\end{abstract}

\section{INTRODUCTION}
As a complex signal, speech carries rich paralinguistic information like speaker trait and speaker state, etc, alongside linguistic content. Recognising emotion and intent from speech has emerged as a crucial task in speech technology. Emotion recognition aims to identify a speaker's emotional states, either as discrete categories (\eg happy, sad, etc) or as continuous values (\ie arousal and valance)~\cite{schuller2013computational}. Intent recognition, on the other hand, focuses on recognising a speaker's goal or purpose, \eg agreeing, wishing, etc~\cite{atuhurra2024domain}. Speech emotion and intent recognition has shown significant potential in various real-world applications, including human-computer interaction, call center conversations, in-car assistance systems, and mental health care~\cite{schuller2013computational,feng2023end,ringeval2019avec,agrawal2023cross}. 

A large amount of speech data nowadays is available for analysis in the aforementioned applications, benefitting from the rapid development of mobile and wearable devices. Yet, data annotation remains labor-intensive, resulting in a lack of labelled data. The scarcity of labelled data has become a bottleneck in improving the model performance for speech emotion and intent recognition~\cite{latif2021survey}.

More recently, end-to-end deep learning models have been proposed and have demonstrated superior performance in speech emotion and intent recognition compared to conventional feature-based models~\cite{ren2023fast,huang2023effective}. These models typically employ complex architectures and leverage self-supervised learning to train on large-scale, unlabelled speech and text datasets~\cite{baevski2020wav2vec,HuBert,Roberta}, thereby exhibiting a strong capability for learning abstract representations. The challenge posed by limited labelled data can be alleviated through the use of representations extracted by pre-trained models. Consequently, the application of pre-trained end-to-end models with transfer learning has become increasingly widespread in speech emotion and intent recognition~\cite{lai2021semi}. 

\begin{figure*}
    \centering
    \includegraphics[width=0.8\linewidth]{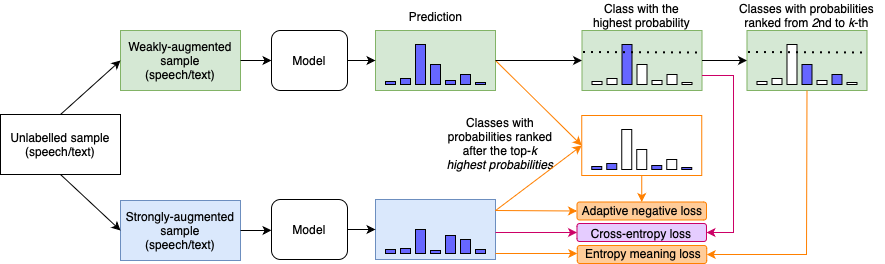}
    \caption{The pipeline of the semi-supervised learning method. The purple lines indicate loss functions in fix-match learning, and the orange ones denote the additional loss functions used in full-match learning.}
    \label{fig:pipeline}
\end{figure*}

In addition to employing end-to-end deep learning models, semi-supervised methods train machine learning models on both labelled and unlabelled data. Several semi-supervised learning approaches have been successfully applied to speech emotion and intent recognition. A type of semi-supervised learning involves applying unsupervised learning to unlabelled data by training an autoencoder, which is then used to extract representations from labelled data for supervised learning~\cite{parthasarathy2020semi,latif2020multi,huang2018speech}. Another type of semi-supervised learning follows an iterative process, including training a model on labelled data using supervised learning, generating pseudo labels for unlabelled data using the trained model, selecting useful pseudo labels, and merging the corresponding unlabelled data and labelled data for further supervised learning~\cite{zhang2016enhanced,feng2022semi,tur2005combining,li2016semi}. A third type of semi-supervised learning directly trains a model on both labelled and unlabelled data with selected pseudo labels. For instance, fix-match learning~\cite{sohn2020fixmatch} has been used to generate pseudo labels derived from video data for unlabelled speech samples~\cite{zhang2021combining}. 


Inspired by the aforementioned third type of semi-supervised learning, we propose applying and comparing two semi-supervised learning approaches, \ie fix-match learning and full-match learning, for speech emotion and intent recognition in a multi-task learning framework. Fix-match learning generates a pseudo label based on the highest predicted probability for an unlabelled sample, whereas full-match learning additionally considers probability distributions across all classes. In both approaches, labelled data undergoes weak augmentation, while unlabelled data is augmented using both weak and strong transformations for better performance. The contributions of this work are manifold. First, a large-scale unlabelled dataset is leveraged in semi-supervised learning to improve model performance beyond what can be achieved with labelled data alone. Second, the semi-supervised learning approaches are comprehensively compared across multiple data augmentation techniques. Third, we apply semi-supervised learning approaches to speech and text modalities, and show that our fused multi-modal models outperform all unimodal models.

\section{METHODOLOGY}

\subsection{End-to-end Speech and Text Foundation Models}
More recently, end-to-end models have been successfully applied to emotion recognition, leveraging pre-trained foundation models through self-supervised learning~\cite{sharma2022multi,ren2023fast}. Given the high complexity of foundation models (\eg HuBERT, which has 1 billion parameters~\cite{HuBert}) and large-scale datasets they are trained on, foundation models have demonstrated superior performance compared to conventional feature-based models~\cite{chang2023knowledge}. We employ the pre-trained HuBERT~\cite{HuBert} and RoBERTa~\cite{Roberta} models to process the speech data and transcripts, respectively. To predict the emotion and intent classes, we introduce two linear layers following the final respective transformer layers. Each linear layer is dedicated to one of the two classification tasks, enabling the models to be trained via multi-task learning.

\subsection{Semi-supervised learning}
Unlike supervised learning, where the labelled data and their corresponding labels $\{\bm{x_l}, y_l\}$ are used to train a model $f$, semi-supervised learning employs unlabelled data $\bm{x_u}$ to further improve the model performance. In this section, we will introduce the \emph{fix-match learning} and \emph{full-match learning} training strategies for utilising unlabelled data (see Fig.~\ref{fig:pipeline}), and explore their application in multi-task learning. 

\subsubsection{Fix-match learning}
For training on labelled data, the labelled samples are weakly augmented as $\bm{x^w_l}$, and then a cross-entropy loss is applied. Since applying cross-entropy loss to unlabelled data is challenging, fix-match learning aims to generate pseudo labels for the unlabelled samples. For this purpose, the unlabelled data is firstly augmented using weak and strong transformations, leading to $\bm{x^w_u}$ and $\bm{x^s_u}$, respectively. The strong augmentation applies more changes to the data compared to the weak augmentation. We assume the predicted class distributions based on the weakly augmented data as $f(\bm{x^w_u})$. The corresponding predicted classes $\mbox{argmax}f(\bm{x^w_u})$ are used as the pseudo labels when their predicted probability exceeds a threshold, \ie $\max f(\bm{x^w_u})>\tau$. Hence, the model is trained based on the loss function 
\begin{equation}
\mathcal{L}_{fix}=\mathcal{L}_c(x^w_l,y_l)+\lambda_1\mathds{1}(\max f(\bm{x^w_u})>\tau)\mathcal{L}_c(x^s_u,y_u),     
\label{eq:fix}
\end{equation}
where $\mathcal{L}_c$ is the cross-entropy loss function and $\lambda_1$ is a constant factor. 

\subsubsection{Full-match learning}
As fix-match learning only employed the unlabelled samples that the model is confident about, full-match learning incorporates all training samples by adding an adaptive negative loss and an entropy meaning loss, in addition to the fix-match loss. Similar to fix-match learning, the predictions from weakly-augmented data are considered as pseudo labels. We firstly calculate the top-$k$ accuracy, which is determined by counting how often the pseudo labels $f(\bm{x^w_u})$ appear among the top-$k$ highest probabilities in $\bm{f(x^s_u)}$. The value of $k$ is determined when the top-$k$ accuracy is larger than a threshold $\sigma$. The adaptive negative loss aims to minimize the probabilities $\bm{f(x^s_u)}$ which are lower than the top-$k$ probabilities, while the entropy meaning loss reduces the probabilities of all entries within the top-$k$ probabilities, except the highest top-$1$ probability. 

\textbf{Adaptive negative loss.} 
With the selected top-$k$ classes for each unlabelled sample, the adaptive negative loss is used to reduce the probability values of the classes ranked after the top-$k$ ones. This process increases the probability difference between the high-probability classes and the low-probability classes, thereby improving the model's confidence in the top-$k$ classes. The loss function is defined by
\begin{equation}
    \mathcal{L}_a=-\frac{1}{B}\sum^B_{b=1}\sum^C_{c=1}\mathds{1}(Rank(\bm{f(x^w_u)})>k)\log(1-\bm{f(x^s_u)}),
    \label{eq:adapt_neg_loss}
\end{equation}
where $B$ is the batch size and $C$ denotes the number of classes.

\textbf{Entropy meaning loss.} 
In addition to improving the top-$1$ highest probability in fix-match learning and reducing the low probabilities ranked after $k$ in adaptive negative loss, $\mathcal{L}_e$ aims to regulate the probability distribution of the $\{2, ...,k\}$-th classes. These classes are often near decision boundaries, therefore they are confusing for the model. This confusion can result in high probabilities for these classes, ultimately reducing the model's confidence in the top-$1$ class. The entropy meaning loss aims to regulate the $\{2, ...,k\}$-th classes by encouraging them to share similar probabilities, 
\begin{equation}
    y^e_u=\frac{1-\mathds{1}(Rank(\bm{f(x^w_u)})\in [2,k])\bm{f(x^s_u)}}{k-1}.
    \label{eq:softlabel}
\end{equation}
The entropy meaning loss is thereby defined by
\begin{equation}
\begin{split}
    \mathcal{L}_e=-\frac{1}{BC}\sum^B_{b=1}\sum^C_{c=1}\mathds{1}(Rank(\bm{f(x^w_u)})>k)(y^e_u\log\bm{f(x^s_u)} \\
    +(1-y^e_u)\log(1-\bm{f(x^s_u)})).
\end{split}
\label{eq:entropy_mean_loss}
\end{equation}
Finally, the overall loss function is 
\begin{equation}
    \mathcal{L}_{full}=\mathcal{L}_{fix} +\lambda_2 \mathcal{L}_a +\lambda_3 \mathcal{L}_e,
\label{eq:full}
\end{equation} 
where $\lambda_1$ and $\lambda_2$ are constant factors.

\subsubsection{Multi-task Learning}
We apply the semi-supervised learning approaches within a multi-task learning framework, specifically for emotion and intent classification. For an unlabelled sample, the pseudo label in fix-match learning is used only when the model is confident in both emotion and intent predictions, \ie both highest probabilities exceed $\tau$. The fix-match loss for both tasks is defined as
$\mathcal{L}^{all}_{fix}=\mathcal{L}^{emo}_{fix}+\lambda\mathcal{L}^{int}_{fix}$, where $\mathcal{L}^{emo}_{fix}$ and $\mathcal{L}^{int}_{fix}$ are the losses from (\ref{eq:fix}) for emotion recognition and intent recognition, respectively. 
The full-match loss functions are applied to both tasks independently: $\mathcal{L}^{all}_{full}=\mathcal{L}^{emo}_{full}+\lambda\mathcal{L}^{int}_{full}$, where $\mathcal{L}^{emo}_{full}$ and $\mathcal{L}^{int}_{full}$ are the losses from (\ref{eq:full}). 

\section{EXPERIMENTAL RESULTS}

\subsection{Dataset}
We use the English dataset from Track 1 of the MC-EIU database~\cite{liu2024emotion} to evaluate the proposed approach. The dataset comprises dialogue scenarios from various genres, such as comedy, family, and others. We extract the speech samples from the original video files and resample the speech samples into 16\,kHz. In total, the dataset consists of 3,610 labelled speech samples and 46,542 unlabelled speech samples. The labelled data is annotated with seven emotional states (anger, disgust, fear, happy, neutral, sad, and surprise) and eight intent labels (acknowledging, agreeing, consoling, encouraging, neutral, questioning, suggesting, and wishing). Since the labels of the official test set are not publicly available, we re-split the original training set into a new training and validation set, while using the original validation set as our test set. During splitting the original training set, we ensure the label distribution remains consistent between the new training and validation sets. The distribution of the labelled data is given in Table~\ref{tab:data}.

\begin{table}[]
    \centering
    \caption{The distribution of the labelled data across the emotion and intent classes. The training and validation sets are re-split from the original training set.}    
    \vspace{-5pt}    
    \begin{tabular}{l|l|R{.7cm}|R{.7cm}|R{.7cm}|R{.7cm}}
    \toprule
       \textbf{Task} & \textbf{Class} & \textbf{Train} & \textbf{Valid} & \textbf{Test} & \textbf{$\sum$}\\
       \midrule
       \multirow{7}{*}{Emotion}&Anger   & 303 & 58 & 41 & 402 \\
        &Disgust   & 292 & 55 & 59 & 406 \\
        &Fear   & 293 & 56 & 50 & 399 \\
        &Happy   & 355 & 68 & 75 & 498 \\
        &Neutral   & 815 & 155 & 126 & 1,096 \\
        &Sad   & 299 & 57 & 47 & 403 \\
        &Surprise   & 298 & 57 & 51 & 406 \\
        \midrule
        \multirow{8}{*}{Intent}&Acknowledging   & 255 & 58 & 30 & 343 \\
        &Agreeing   & 280 & 50 & 47 & 377 \\
        &Consoling   & 243 & 49 & 56 & 348 \\
        &Encouraging   & 225 & 52 & 37 & 314 \\
        &Neutral   & 666 & 132 & 114 & 912 \\
        &Questioning   & 388 & 80 & 68 & 536 \\
        &Suggesting   & 320 & 50 & 51 & 421 \\
        &Wishing   & 278 & 35 & 46 & 359 \\
        \midrule
        \textbf{$\sum$} & $--$ & 2,655 & 506& 449& 3,610\\
    \bottomrule
    \end{tabular}
    \label{tab:data}
\end{table}

\subsection{Evaluation Metrics}
We evaluate the proposed methods using the F1 score, calculated as the weighted average of class-wise F1 scores, where the weights are determined by the number of true instances for each label. To evaluate the overall model performance across both classification tasks, we follow the MC-EIU challenge to use the Joint Recognition Balance Metrics (JRBM), \ie 
\begin{equation}
    \mbox{JRBM}=\frac{2\cdot F1_{emo}\cdot F1_{intent}}{F1_{emo}+F1_{intent}},
\end{equation}
where $F1_{emo}$ and $F1_{intent}$ are the F1 scores of emotion and intent classifications, respectively.

\subsection{Implementation Details}

\textbf{Automatic speech recognition on unlabelled data.} 
The speech samples from the unlabelled dataset are transcribed into text using OpenAI's `medium.en' Whisper model with 769 million parameters~\cite{radford2022whisper}.

\textbf{Data augmentation.} To compare different augmentation methods, we apply i) three weak augmentations on the speech signals, namely \emph{flipping}, \emph{time masking}, and \emph{pitch shifting}, and ii) a strong augmentation using \emph{Gaussian noise}~\cite{10038009}. Within a speech signal, \emph{flipping} reverses a randomly selected segment with a maximum duration of 6.25\,s, \emph{time masking} randomly sets a sub-clip (up to 30k frames) to zero, and \emph{pitch shifting} alters the speech pitch within 4 steps. \emph{Gaussian noise} is added to the entire speech signal with a scaling factor of 0.05. We also employ i) three weak augmentations on the transcripts, \ie \emph{swapping}, \emph{deleting}, and \emph{synonym}, and ii) a strong augmentation of \emph{contextual}~\cite{wei2019eda}. \emph{Swapping} randomly exchanges adjacent words, \emph{deleting} randomly removes partial words, and \emph{synonym} substitute words with semantically similar alternatives. \emph{Contextual} identifies the top $n$ similar words using contextual word embeddings for replacement. All hyperparameters are set experimentally.

\textbf{Model training.} The models are trained using the Adam optimizer with an initial learning rate of $3\times 10^{-5}$. The learning rate is reduced by a factor of 0.9 at each epoch. The batch size is set to 8 for speech models trained with fix-match learning and full-match learning due to high memory requirements, while a batch size of 16 is used for all other models. The loss function coefficients (\ref{eq:fix}) and (\ref{eq:full}) are set as $\lambda_1=\lambda_2=\lambda_3=0.5$, with $\lambda$ set to 1. The hyperparameters are set as $\tau=0.95$ and $\sigma=0.99$.

\subsection{Results and Discussion}

\begin{table*}[]
    \centering
    \caption{The performance in terms of F1 score and JRBM achieved by the HuBERT and RoBERTa models on speech and text modalities, respectively. The models are compared in the settings with and without weak augmentation on the unlabelled data. Different weak augmentation methods are also compared.}   
    \vspace{-2pt}    
    \begin{tabular}{l|c|c|P{1.1cm}|P{1.1cm}P{1.1cm}P{1.1cm}|P{1.1cm}|P{1.1cm}P{1.1cm}P{1.1cm}}
    \toprule
            \multirow{3}{*}{\textbf{Model}}&\multirow{3}{*}{\textbf{Method}}& \multirow{3}{*}{\textbf{Augment}}&\multicolumn{4}{c|}{\textbf{wo/ weak augmentation}} &\multicolumn{4}{c}{\textbf{w/ weak augmentation}} \\
            \cline{4-11}
          &  & & \textbf{Valid} &\multicolumn{3}{c|}{\textbf{Test}} &\textbf{Valid} &\multicolumn{3}{c}{\textbf{Test}} \\
         \cline{4-11}
          && & \textit{JRBM} & \textit{F1 emo} & \textit{F1 intent} & \textit{JRBM} & \textit{JRBM} & \textit{F1 emo} & \textit{F1 intent} & \textit{JRBM} \\
         \midrule
         \multirow{7}{*}{HuBERT} &Baseline& -- & 0.430 & 0.292 & 0.323 & 0.307 & 0.430 & 0.292 & 0.323 & 0.307\\
          &Fix-match& Flipping & 0.386 & 0.271 & 0.266 & 0.268 & 0.436 & \textbf{0.301} & \textbf{0.399} & \textbf{0.343}\\
          &Fix-match& Time masking & 0.364 & 0.260 & 0.272 & 0.266 & 0.438 &0.298 & 0.300 & 0.299\\
          &Fix-match& Pitch shifting& 0.393 & 0.297 & 0.319 & 0.308 & 0.377 & 0.316 & 0.334 & 0.324\\
         
          &Full-match& Flipping & \textbf{0.434} & 0.317 & 0.320 & 0.318 & \textbf{0.444} & 0.280 & 0.339 & 0.307\\
          &Full-match& Time masking &  0.426 & 0.262 & 0.343 & 0.297 & 0.361 & 0.267 & 0.334 & 0.296\\
          &Full-match& Pitch shifting & 0.404 & \textbf{0.317} & \textbf{0.353} & \textbf{0.334} & 0.392 & 0.287 & 0.333 & 0.308\\
          \midrule
         \multirow{7}{*}{RoBERTa} &Baseline& -- & 0.474& 0.268 & 0.417 & 0.326 & 0.474& 0.268 & 0.417 & 0.326 \\
          &Fix-match& Swapping & \textbf{0.496} & 0.292 & 0.407 & 0.340 & 0.447 & 0.306 & 0.411 & 0.351\\
          &Fix-match& Deleting & 0.481 & 0.315 & 0.420 & 0.360 & 0.464 & 0.305 & 0.406 & 0.348\\
          &Fix-match& Synonym & 0.444 & 0.289 & 0.429 & 0.345 & 0.454 & \textbf{0.338} & \textbf{0.453} & \textbf{0.387}\\
          &Full-match& Swapping & 0.468 & 0.290 & 0.438 & 0.349 & 0.470 & 0.311 & 0.436 & 0.363\\
          &Full-match& Deleting & 0.468 & 0.301 & 0.437 & 0.356 & \textbf{0.475} & 0.314 & 0.405 & 0.354 \\
          &Full-match& Synonym & 0.475 & \textbf{0.306} & \textbf{0.443} & \textbf{0.362} & 0.473 & 0.309 & 0.440 & 0.363\\
         \midrule
         Best 2 & Fusion & -- & 0.526 & \textbf{0.351} & \textbf{0.454} & \textbf{0.396} & \textbf{0.529} & \textbf{0.370} & \textbf{0.512} & \textbf{0.430}   \\
         Best 4 & Fusion & -- & \textbf{0.528} & 0.347 & 0.435 & 0.386 & 0.513 & 0.345 & 0.484 & 0.403\\
         \bottomrule
    \end{tabular}
    \label{tab:result}
\end{table*}
The chance levels of the emotion and intent classification tasks are 0.143 and 0.125, respectively.
For the baselines, we fine-tune the pre-trained HuBERT model for processing the speech data, and the pre-trained RoBERTa model for making predictions from the transcriptions, as shown in Table~\ref{tab:result}. The results indicate that RoBERTa performs better than HuBERT on both validation and test sets. This might be caused by noise in the speech files, such as laugh tracks, as well as variations in speaker accents. When comparing the semi-supervised learning approaches with the baselines, both fix-match learning and full-match learning mostly perform better than the baselines, demonstrating that incorporating unlabelled data helps improve model performance.

\textbf{Ablation study.}
To assess the impact of data augmentation on unlabelled data, we evaluate the effect of weak augmentation on model performance. In Table~\ref{tab:result}, models trained with weak augmentation mostly generally outperform those without it. Full-match learning performs better than fix-match learning in the absence of weak augmentation, whereas fix-match learning yields superior results when weak augmentation is applied. This might be because weak augmentation increases the complexity of the data distribution, making it more challenging to train models with multiple loss components in full-match learning. 

Additionally, we compare different weak augmentation methods across both semi-supervised learning methods. In the speech modality, \emph{flipping} and \emph{pitch shifting} outperform \emph{time masking}. A possible reason is that \emph{time masking} removes partial speech frames, and the duration of the masked speech frames might be not optimal. Similarly, in the text modality, \emph{deleting} performs slightly worse than the other two weak augmentation methods. \emph{Synonym} performs the best, probably because it enriches the vocabulary within the overall texts.

\textbf{Model fusion.}
We combine the best-performing models from HuBERT and RoBERTa using a late fusion (see `best 2' in Table~\ref{tab:result}), as well as the top two models from each modality (see `best 4' in Table~\ref{tab:result}). The late fusion herein is margin sampling value~\cite{ren2018deep}, which selects the predicted label based on the largest difference between the highest and second highest probabilities. The `best 2' fusion performs better than `best 4'.
The `best 2' fusion with a JRBM score of 0.430 outperforms all unimodal models, suggesting that HuBERT and RoBERTa probably capture complementary information from the data. 

\begin{figure}
    \centering
    \includegraphics[width=0.7\linewidth]{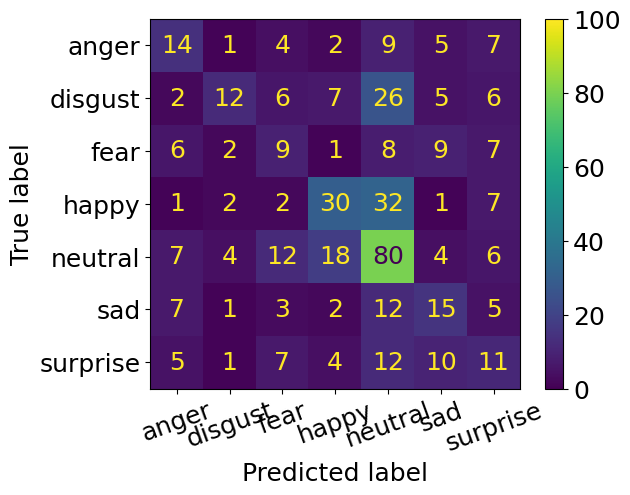}
    \vspace{-5pt}
    \caption{Confusion matrix of fusing the best 2 models for emotion recognition on the test set.}
    \label{fig:confmat_emo}
\end{figure}

\begin{figure}
    \centering
    \includegraphics[width=0.99\linewidth]{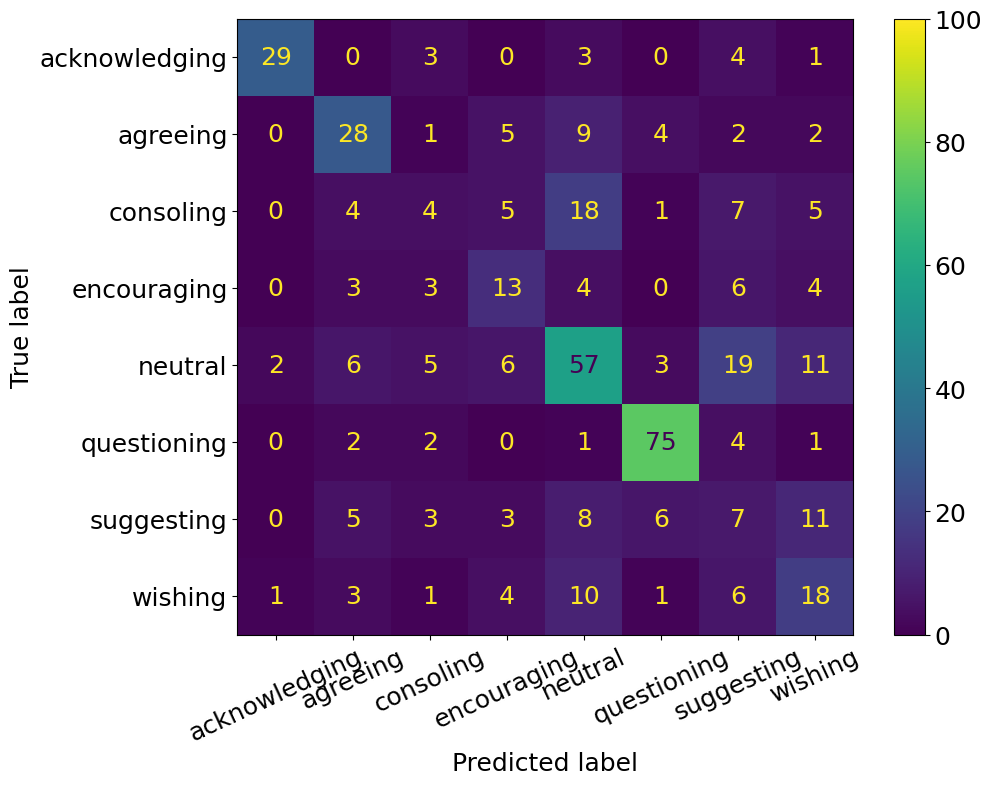}
    \vspace{-5pt}
    \caption{Confusion matrix of fusing the best 2 models for intent recognition on the test set.}
    \label{fig:confmat_int}
\end{figure}

Finally, the confusion matrices of the best-performing model, \ie the `best 2' fusion, for emotion and intent recognition are presented in Fig.~\ref{fig:confmat_emo} and Fig.~\ref{fig:confmat_int}, respectively. In Fig.~\ref{fig:confmat_emo}, 
`neutral' is the preferred class
probably because of class imbalance. 
Partial `Fear' samples are predicted into `sad', probably caused by their high similarities in high pitch (`sad' with crying despair), falsetto voice (`sad' with crying despair), and breathiness~\cite{burkhardt2000verification,yuan2002acoustic}. Similarly, some `surprise' samples are misclassified as `sad'.
In Fig.~\ref{fig:confmat_int}, the `questioning' class is predicted the best. The reason might be some `questioning' sentences (\eg yes or no) are spoken with a rising intonation, making them acoustically distinct from other intent categories.

\section{CONCLUSIONS AND FUTURE WORK}
We conducted end-to-end speech emotion and intent recognition models in multi-task learning. To enhance model performance, we introduced two semi-supervised learning approaches, \ie fix-match learning and full match learning. The impact of weak augmentation and different augmentation approaches were thoroughly investigated in an ablation study. Ultimately. the fused best two models achieved the highest performance in our experiments. In future efforts, additional end-to-end models will be explored, \eg WavLM~\cite{chen2022wavlm}. The hyperparameter settings, including loss coefficients and augmentation parameters, can be further optimised to improve the performance. Moreover, one can explore annotating the unlabelled data with the help of additional models, such as large language models~\cite{latif2023can}.








\balance

\bibliographystyle{IEEEtran}
\bibliography{ref}

\end{document}